\documentclass[preprint,pra,aps,superscriptaddress,showpacs,prd]{revtex4}
\usepackage{epsfig,amsmath,graphicx,amssymb}
\def\be{\begin{equation}}
\def\ee{\end{equation}}
\def\bea{\begin{eqnarray}}
\def\eea{\end{eqnarray}}
\begin{document}
\title{Electrostatic and Magnetostatic Solutions
in a Lorentz-Violating Electrodynamics Model
 }
\author{Jianfeng Wu } \address{Department
of Physics and Institute of Theoretical Physics, East China Normal
University, Shanghai, China, 200062}
\author{Xun Xue }\email{xxue@phy.ecnu.edu.cn} \address{Department of Physics and
Institute of Theoretical Physics, East China Normal University,
Shanghai, China, 200062}

\date{\today}

\begin{abstract}

We propose an effective Lorentz violating electrodynamical model
via static de Sitter metric which is deviated from Minkowski
metric by a minuscule amount depending on the cosmological
constant. We obtain the electromagnetic field equations via the
vierbein decomposition of the tensors. In addition, as an
application of the electromagnetic field equations obtained, we
get the solutions of electrostatic field and magnetostatic field
due to a point charge and a circle current respectively and
discussed the implication of the effect of Lorentz violation in
our electromagnetic theory.


\end{abstract}

\maketitle

Lorentz invariance is one of the most great discoveries of physics
in the history of physics and has been confirmed to ever greater
precision. Most of the evidence comes from short distance tests.
However there are plenty of signs that something strange may
happen at large distance (for example,dark energy), where the
constraints on Lorentz symmetry are much weaker. It is reasonable
that many researchers are interesting in Lorentz violation(LV)
from various points of views \cite{mew,jac,col,leh,blu,bm}. They
pointed out that Lorentz invariance can be viewed as a low energy
effective invariance. Remarkably, under suitable circumstances,
some experimental information about quantum gravity can
nonetheless be obtained. The point is that minuscule effects
emerging from the underlying quantum gravity might be detected in
sufficiently sensitive experiments. To be identified as definitive
signals from the Planck scale, such effects would need to violate
some established principle of low-energy physics. One promising
class of potential effects is relativity violations, arising from
breaking the Lorentz symmetry that lies at the heart of
relativity.  Recent proposals suggest LV effects could emerge from
strings, loop quantum gravity, noncommutative field theories, or
numerous other sources at the Planck scale\cite{JLM}. On the other
hand, recent observations, such as the luminosity observations of
the farthest supernovas \cite{per}, show that our universe is
accelerated expanding , asymptotic de Sitter with a positive
cosmological constant $\Lambda$ \cite{rie,slu,ben}.

Among the developments on LV research, one is a systematic
extension of the standard model of particle physics incorporating
all possible LV in the renormalizable sector called Standard Model
Extension(SME), developed by Colladay and
Kosteleck\'{y}~\cite{CK}. That provided a framework for computing
in effective field theory the observable consequences for many
experiments and led to much experimental work setting limits on
the LV parameters in the Lagrangian~\cite{AKbook}.

We limit our attention in the present work to the sector of
classical Lorentz violating electrodynamics in Minkowski
space-time , coupled to an arbitrary 4-current source, in this
framework of SME. Due to the presence of dark energy or the
nonzero positive cosmological constant, the space-time without any
matter is de Sitter rather than Minkowskian and so it is a natural
way to substitute the Lorentz invariant low energy effective
theory with its covariant formulation in de Sitter space-time for
a field theory in Minkowski space-time. The Lorentz invariance is
violated obviously in this way to the observer in Minkowski
space-time. We mean in this approach that the Lorentz symmetry is
an approximately symmetry of the low energy effective theory.
There are different types of metric for de Sitter space-time and
it is well known that quantum field theory in de Sitter space-time
equipped with the static metric is a finite temperature field
theory in a pure field theory in curved space-time
approach\cite{SSV}. However, we investigate the Lorentz violating
electrodynamics in Minkowski space-time in the present works and
so there is no finite temperature problem here. In this way,
different choices of de Sitter metrics to formulate the Lorentz
violating theory is just following different scenarios to violate
Lorentz symmetry. In this paper we propose a scenario that the
substitute of Lorentz invariant electromagnetic theory is its
formulation in de Sitter space-time equipped with the static
metric which is obviously Lorentz violating. We define all the
observables in  Minkowski space-time as the corresponding vierbein
decomposition components in de Sitter space-time of corresponding
physical quantities.

We now set up the LV electrodynamics by the vierbein formalism. A
basic object in the formalism is the vierbeins
$\vartheta^{a}_{\mu}$, which can be viewed as providing at each
point on the space-time manifold a link between the covariant
components $T_{\lambda\mu\nu\ldots}$ of a tensor field in a
coordinate basis and the corresponding covariant components
$T_{abc\ldots}$ of the tensor field in a local Lorentz frame. The
link is given by
\be
T_{\lambda\mu\nu\ldots}=\vartheta^{a}_{\,\,\lambda}\vartheta^{b}_{\,\,\mu}\vartheta^{a}_{\,\,\nu}\ldots
T_{abc\ldots}
\ee
In the coordinate basis, the components of the space-time metric
are denoted $g_{\mu\nu}$. In the local Lorentz frame, the metric
components take the Minkowski form $\eta_{ab}=diag(1,-1,-1,-1)$.

Then we'd like to introduce the de Sitter space and its metric, de
Sitter space can be regarded as a 4-d hyperboloid $S_{R}$ embedded
in a 5-d Minkowski space with $\eta_{AB}=diag(1,-1,-1,-1,-1)$,
\bea
S_{R}:\,\,\,\,\eta_{AB}\xi^{A}\xi^{B}=-R^{2}, \nonumber\\
ds^{2}=\eta_{AB}d\xi^{A}d\xi^{B},
\eea
where $A,B=0,...,4$. Clearly, Eqs. (2) are invariant under de
Sitter group $SO(1,4)$. The metric of this space-time can be
written as \cite{wei}
\be
ds^2  = \eta _{\mu \nu } d\xi ^\mu  d\xi ^\nu   - \frac{{K(\eta
_{\mu \nu } \xi ^\mu  d\xi ^\nu  )^2 }}{{1 + K(\eta _{\mu \nu }
\xi ^\mu  \xi ^\nu  )}}
\ee
where $\mu,\nu=0,...,3$, $K=\frac{1}{R^{2}}=\frac{\Lambda}{3}$,
$\Lambda$ is the cosmological constant. This metric is invariant
under two classes of simple transformations (see, for example, P
387 of the book \cite{wei}), one is $SO(1,3)$ transformations:
\be
\xi ^{'\mu }  = L_{\,\,\,\nu} ^\mu  \xi ^\nu
\ee
the other is the 'quasitranslations', with
\bea
\xi ^{'\mu }  = \xi ^\mu   + a^\mu  [(1 - K\eta _{\rho \sigma }
\xi ^\rho  \xi ^\sigma  )^{1/2}  - bK\eta _{\rho \sigma } \xi
^\rho  a^\sigma  ]\\
b=\frac{{1 - (1 - K\eta _{\rho \sigma } a^\rho  a^\sigma )^{1/2}
}}{{K\eta _{\rho \sigma } a^\rho  a^\sigma  }} \nonumber
\eea
In particular, these transformations take the origin $\xi^{\mu}=0$
into any $a^{\mu}$. For the metric given by Eq. (3), we can
introduce coordinates in which the metric appears time-independent
by
\bea
x^{i}=\xi^{i}=x'\,^{i} \exp (K^{1/2} t'),\\ \nonumber
 \xi ^0  =
\frac{1}{{\sqrt K }}[\frac{{K\textbf{x}'^2 }}{2}\cosh (K^{1/2} t')
+ (1 + \frac{{K\textbf{x}'^2 }}{2})\sinh (K^{1/2} t')],\\
\nonumber t=t'-
\frac{1}{{2K^{1/2} }}\ln [1 - K\textbf{x}'^2 \exp (2K^{1/2} t')].\\
\nonumber
\eea
Then Eq. (3) becomes
\be
d s^{2}=(1-K \textbf{x}^{2})d
t^{2}-d\textbf{x}^{2}-\frac{K(\textbf{x}\cdot
d\textbf{x})^{2}}{1-K \textbf{x}^{2}}.
\ee
It can be showed that the spatial metric of the space-time is just
the metric of a 3-d spherical surface (with radius $R$) in 4-d
Euclidean space. However, unlike the metric given by Eq. (3), this
static de Sitter metric is obviously Lorentz violating. Noting
that the transformation (5) and (6) leaving the metric (7)
invariant also take the spatial origin $\textbf{x}=\textbf{0}$
into any $\textbf{a}$ and contain the spatial $SO(3)$ rotation.
Choosing the spherical coordinate, we can rewrite the static
metric Eq. (7) as follow
\be
ds^{2}=\sigma dt^{2} - \frac{1}{\sigma}dr^{2} -
r^{2}d\theta^{2}-r^{2}\sin\theta ^{2} d\phi ^{2},
\ee
where $\sigma=1-Kr^{2}$. So we can define a local Lorentz frame
with vierbeins $\vartheta^{a}_{\mu},a=0,1,2,3$, where
\be
\vartheta^{a}_{\mu}=diag(\sqrt{\sigma}, \frac{1}{\sqrt{\sigma}},
r, r \sin{\theta}).
\ee

As in general relativity, the observables are vectors and tensors
in the local Lorentz frame. Here we define the observables in
Minkowski space-time as the vierbein decomposition components of
the corresponding tensors of physical quantities in de Sitter
space-time. In the present work, what we concerned is that the
observables of electromagnetic field, the electric field strength
$\overrightarrow{\textbf{E}}$ and magnetic field strength
$\overrightarrow{\textbf{B}}$.

First, we introduce the electromagnetic potential contravariant
vector
\be
A^{\mu}= e^{\mu}_{a} A^{a}=(\frac{1}{\sqrt{\sigma}}\varphi,
\sqrt{\sigma}A_{r}, \frac{1}{r} A_{\theta}, \frac{1}{r
\sin{\theta}}A_{\phi})
\ee
where
\be
e^{\mu}_{a}=\eta_{ab}g^{\mu\nu}\vartheta^{b}_{\nu}=
diag(\frac{1}{\sqrt{\sigma}}, \sqrt{\sigma},\frac{1}{r},
\frac{1}{r \sin{\theta}}), A^{a}=(\varphi, A_{r}, A_{\theta},
A_{\phi})\nonumber
\ee
and $A^{a}$ are components of the 'ordinary' vector \cite{wei}.
This vector is that what we are seeking for, i.e. the observable
vector. So the covariant 1-form can be written as below(we define
$x^{\mu}=(t, r, \theta, \phi)$ hereafter),
\be\label{aaa}
A=A_{\mu}d x^{\mu}=\sqrt{\sigma}\varphi dt
-\frac{1}{\sqrt{\sigma}}A_{r}dr - r A_{\theta}d \theta - r
\sin{\theta}A_{\phi}d \phi.
\ee
Then, we can introduce the electromagnetic field-strength
covariant 2-form $F=dA=\frac{1}{2}F_{\mu\nu}dx^{\mu}\wedge
dx^{\nu}$. Accordingly, here $F_{ab}$ are components of the
'ordinary' electromagnetic field strength tensor and they can be
written down as a matrix:
\be\label{ccc}
F_{ab}=\left(%
\begin{array}{cccc}
  0 & -E_{r} & -E_{\theta} & -E_{\phi} \\
  E_{r} & 0 & B_{\phi} & -B_{\theta} \\
  E_{\theta} & -B_{\phi} & 0 & B_{r} \\
  E_{\phi} & B_{\theta} & -B_{r}& 0 \\
\end{array}%
\right)\vspace{-4mm}
\ee
Then $F_{\mu\nu}$ becomes
\be\label{ddd}
F_{\mu\nu}=\left(%
\begin{array}{cccc}
  0 & -E_{r} & -r \sqrt{\sigma}E_{\theta} & -r \sin{\theta}\sqrt{\sigma}E_{\phi} \\
  E_{r} & 0 & \frac{r}{\sqrt{\sigma}}B_{\phi} & -\frac{r \sin{\theta}}{\sqrt{\sigma}}B_{\theta} \\
  r \sqrt{\sigma}E_{\theta} & -\frac{r}{\sqrt{\sigma}}B_{\phi} & 0 & r^{2}\sin{\theta}B_{r} \\
  r \sin{\theta}\sqrt{\sigma}E_{\phi} & \frac{r \sin{\theta}}{\sqrt{\sigma}}B_{\theta} & -r^{2}\sin{\theta}B_{r}& 0 \\
\end{array}%
\right).
\ee
The action of the electromagenetic field can be writen as
\be
I_{M}=\int (-F\wedge \ast F-A\wedge\ast j).
\ee
Thereinto, the symbol '$\ast$' is the Hodge-dual operator. As
doing with the familiar formulae for gradient, curl, and
divergence in the classical curvilinear coordinate systems, we now
introduce these things in the spatial part of the local Lorentz
frame, or accurately, on the submanifold $S^{3}$ of static de
Sitter space-time manifold, by
\be
\stackrel{\sim}\nabla\psi=d'\psi=\sqrt{\sigma}\psi
_{,\,\,r}\vartheta^{1} + \frac{1}{r}\psi
_{,\,\,\theta}\vartheta^{2} + \frac{1}{r\sin{\theta}}\psi
_{,\,\,\phi}\vartheta^{3},
\ee
\be
\stackrel{\sim}\nabla\cdot \vec{f}=\ast ' d'\ast '
f=\frac{2}{r}\sqrt{\sigma}f_{r}+ \sqrt{\sigma}f_{r,\,\,r} +
\frac{\cos{\theta}}{r\sin{\theta}}f_{\theta} +\frac{1}{r}
f_{\theta,\,\,\theta} + \frac{1}{r\sin{\theta}}f_{\phi,\,\,\phi},
\ee
\bea
\stackrel{\sim}\nabla\times \vec{f}=\ast ' d'
f=(\frac{\cos{\theta}}{r\sin{\theta}}f_{\phi}+
\frac{1}{r}f_{\phi,\,\,\theta} - \frac{1}{r\sin{\theta}}f_{\theta
,\,\,\phi})\vartheta^{1}  \\ \nonumber+
(\frac{1}{r\sin{\theta}}f_{r,\,\,\phi} -
\sqrt{\sigma}f_{\phi,\,\,r} -\frac{\sqrt{\sigma}}{r}
f_{\phi})\vartheta^{2}
\\
\nonumber+(\frac{\sqrt{\sigma}}{r}f_{\theta}+\sqrt{\sigma}f_{\theta,\,\,r}-\frac{1}{r}f_{r,\,\,\theta})\vartheta^{3},
\eea
\bea
\stackrel{\sim}\nabla^{2} \psi=(d'\delta'+\delta'
d')\psi=\frac{\sqrt{\sigma}}{r^{2}}\frac{\partial}{\partial
r}(r^{2}\sqrt{\sigma}
\psi_{,\,\,r})+\frac{1}{r^{2}\sin{\theta}}\frac{\partial}{\partial
\theta}(\sin{\theta}\psi_{,\,\,\theta})+\frac{1}{r^{2}\sin^{2}{\theta}}\frac{\partial^{2}\psi}{\partial
\phi^{2}},
\eea
where $\vec{f}$ is an 'ordinary' 3-d vector on $S^{3}$ and
$\vec{f}=f_{r}\vartheta
^{1}+f_{\theta}\vartheta^{2}+f_{\phi}\vartheta^{3}=(f_{r},f_{\theta},f_{\phi})$.
The independent basis $\vartheta^{i}=\vartheta^{i}_{\mu}d
x^{\mu},\,\,i=1,2,3$ point out the orient of a 3-d vector in the
local Lorentz frame. We use the symbol tilde and prime to indicate
that we do these things on the submanifold $S^{3}$.

Noticing that Eq.(\ref{aaa})-(\ref{ddd}) build a bridge from
$A^{a}$ to $\overrightarrow{\textbf{E}}$ and
$\overrightarrow{\textbf{B}}$, we can write down these relational
equations to show it clearly, by performing some elementary
calculations,
\be
\vec{E}=-\frac{1}{\sqrt{\sigma}}
\stackrel{\sim}\nabla(\sqrt{\sigma}\varphi)
-\frac{1}{\sqrt{\sigma}}\frac{\partial \vec{A}}{\partial t}
\ee
and
\be
\vec{B}= \stackrel{\sim}\nabla \times \vec{A}
\ee
where $\vec{A}=(A_{r},A_{\theta},A_{\phi})$,
$\vec{E}=(E_{r},E_{\theta},E_{\phi})$ and
$\vec{B}=(B_{r},B_{\theta},B_{\phi})$, respectively.

Noticing that $F=dA$ is an exact 2-form, we can immediately obtain
the Bianchi identity $ dF=d^{2}A\equiv 0$ and the dynamical
equation $\delta F= \ast d \ast F= j= j_{\mu}dx^{\mu}$, where we
define
\be
j^{\mu}=\vartheta^{\mu}_{a}j^{a}=(\frac{1}{\sqrt{\sigma}}\rho,
\sqrt{\sigma}j_{r}, \frac{1}{r}j_{\theta}, \frac{1}{r
\sin{\theta}} j_{\phi})
\ee
with $j_{a}=(\rho, j_{r}, j_{\theta}, j_{\phi})$. The 'ordinary'
electric current density can be defined as before
$\vec{j}=(j_{r},j_{\theta},j_{\phi})$ . Then the electromagnetic
field equations in static de Sitter space-time are easy to
obtained as bellow
\be
\stackrel{\sim}\nabla\cdot\vec{B}=0
\ee
\be
\stackrel{\sim}\nabla\times (\sqrt{\sigma}\vec{E}) +
\frac{\partial \vec{B}}{\partial t} = 0
\ee
\be
\stackrel{\sim}\nabla\cdot\vec{E}=\rho
\ee
\be
\stackrel{\sim}\nabla\times\vec{B}-\frac{1}{\sqrt{\sigma}}\frac{\partial\vec{E}}{\partial
t}=\vec{j}
\ee

In addition, we study the covariant gauge condition of the
electromagnetic field in static de Sitter space-time. In static de
Sitter space-time, the reasonable gauge condition is the de Sitter
covariant gauge condition $\delta A=0$, in the local Lorentz
frame, one can write this equation as follows:
\be
\stackrel{\sim}\nabla\cdot(\sqrt{\sigma}\vec{A})+\frac{\partial
\varphi}{\partial t}=0
\ee
This de Sitter gauge will play an important role in dealing with
the magnetostatic field.

Now we would like to investigate the interaction between the
electromagnetic field and a charged source $j^{\mu}$ in static de
Sitter space-time, as we do in Lorentz invariant electrodynamics,
we write the Lagrange density of the system reads
\be
\mathcal{L}_{\mathcal{M}}=\mathcal{L}_{E}+\mathcal{L'}_{\mathcal{M}}=-\frac{1}{4}F^{\mu\nu}F_{\mu\nu}-j_{\mu}A^{\mu}
\ee
here $\mathcal{L}_{E}=-\frac{1}{4}F^{\mu\nu}F_{\mu\nu}$ is the
purely electromagnetic term and another term
$\mathcal{L}_{\mathcal{M}}$ describes the charged particles (with
charge $e$) and their electromagnetic interactions. Then the
electromagnetic force $f^{\mu}(x)$ can be obtained as
\be
f^{\mu}=F^{\mu}_{\,\,\,\gamma}j^{\gamma}=(e\frac{\vec{v}\cdot\vec{E}}{\sqrt{\sigma}},
\sqrt{\sigma}f_{r},\frac{1}{r}f_{\theta},\frac{1}{r\sin{\theta}}f_{\phi})
\ee
with
\be
\vec{f}=(f_{r},f_{\theta},f_{\phi})=e(\vec{E}+\vec{v}\times\vec{B}).
\nonumber
\ee
If $K\rightarrow0$,  the LV electrodynamics tends to Lorentz
invariant one and the force returned to Lorentz force. If we
defined the purely electromagnetic term of energy-momentum tensor
as usual,
\be
T_{em}^{\alpha\beta}\equiv
F^{\alpha}_{\,\,\,\gamma}F^{\alpha\beta}-\frac{1}{4}g^{\alpha\beta}F_{\lambda\delta}F^{\lambda\delta},
\ee
we can obtain the energy-momentum conservation law for LV
electrodynamics.
\be
T_{em\,\, \textbf{;}\,\,\beta}^{\alpha\beta}=
-F^{\alpha}_{\,\,\,\beta}j^{\beta}=-f^{\alpha}.
\ee
The symbol semicolon is the abbreviation for covariant derivative.
To make this tensor equation familiar to us and have obviously
observable meaning, we should rewrite the equation using the
vierbein formalism as
\be
\frac{1}{\sqrt{\sigma}}\stackrel{\sim}\nabla\cdot\vec{S}+\frac{1}{\sqrt{\sigma}}\frac{\partial
\omega}{\partial t}=- \vec{j}\cdot\vec{E},
\ee
\be
\vec{S}=\sigma (\vec{E}\times\vec{B}), \omega =
\frac{1}{2}(E^{2}+B^{2}) ,\nonumber
\ee
\be
\stackrel{\sim}\nabla\cdot\stackrel{\rightharpoonup
\rightharpoonup}{\mathcal{J}}+\frac{1}{\sqrt{\sigma}}\frac{\partial
\vec{g}}{\partial t}- K(\vec{r}\times\vec{E}\times\vec{E})=-
\vec{f},
\ee
and
\be
\stackrel{\rightharpoonup
\rightharpoonup}{\mathcal{J}}=-\vec{E}\vec{E}-\vec{B}\vec{B}+\frac{1}{2}\stackrel{\rightharpoonup
\rightharpoonup}{\mathcal{I}}(E^{2}+B^{2}),\vec{g}=\vec{E}\times\vec{B},
\nonumber
\ee
where $\vec{S}$, $\omega$, $\stackrel{\rightharpoonup
\rightharpoonup}{\mathcal{J}}$, and $\vec{g}$ are the energy flux
density (Poynting vector), the energy density the electromagnetic
stress tensor and the momentum density of the system,
respectively. $\stackrel{\rightharpoonup
\rightharpoonup}{\mathcal{I}}$  is the unit tensor in static de
Sitter space-time. Eq. (31) and Eq. (32) are the vierbein
formalism of energy-momentum conservation law  of the LV
electrodynamics. One can observe again that these equations are
different from their cousins in Lorentz invariant formalism.

As an application of the electromagnetic field equations in static
de Sitter space-time, we now introduce the electrostatic field due
to a point charged particle. Let's pay a little attention on a
point charge (with charge $q$) at the point
$\vec{r}_{0}=(r_{0},\theta_{0},\phi_{0})$. One may ask a question:
how to define a real point charge in the local frame? The answer
is connected with the current conservation law in static de Sitter
space-time, that is $\delta j=0$. Using the vierbein formalism,
this conservation law can be written as
\be
\stackrel{\sim}\nabla\cdot(\sqrt{\sigma}\vec{j})+\frac{\partial
\rho}{\partial t}=0.
\ee
Rewriting this equation in the form of spherical coordinate, it is
\be
\nabla\cdot\vec{\stackrel{\sim}j}+\frac{1}{\sqrt{\sigma}}\frac{\partial
\rho}{\partial t}=0,
\ee
where
$\vec{\stackrel{\sim}j}=(\sqrt{\sigma}j_{r},j_{\theta},j_{\phi})$.
The delta function at the point $\vec{r_{0}}$ in $S^{3}$ can be
writen as $\delta
'\,^{3}(\vec{r}-\vec{r_{0}})=\sqrt{\sigma}\delta^{3}(\vec{r}-\vec{r_{0}})$,
so now we can define the charge distribution function of a point
charge as $\rho=q\delta '\,^{3}(\vec{r}-\vec{r}_{0})$. Using the
field equations we obtained above, then the electrostatic field
equation becomes
\be
-\stackrel{\sim}\nabla\cdot
(\frac{1}{\sqrt{\sigma}}\stackrel{\sim}\nabla
(\sqrt{\sigma}\varphi))=q\delta '\,^{3}(\vec{r}-\vec{r_{0}}).
\ee
Utilizing the spherical coordinate, we can transform the equation
to appear the formalism what we are familiar with,
\be
-\nabla^{2}\varphi+K\frac{\partial}{\partial
r}(r^{2}\frac{\partial\varphi}{\partial r})+3K\varphi+2K
r\frac{\partial\varphi}{\partial r}+
\frac{K^{2}r^{2}}{\sigma}=q\delta'\,^{3}(\vec{r}-\vec{r_{0}})
\ee

However, this equation is not easy to solve. Fortunately, there is
a way to round this difficulty, because the equation above is de
Sitter invariance, so one can perform a suitable
'quasitranslation' in static de Sitter space-time to take the
spatial origin $\textbf{x}=\textbf{0}$ into any $\textbf{a}$. Then
we can always choose the observed point as the origin of the local
frame, namely let $r\rightarrow 0$ in Eq. (36). We then arrive at
\be
 -\nabla ^{2}\varphi+ 3K \varphi=q\delta
 ^{3}(\vec{r}-\vec{r}_{0}).
\ee
This equation is very easy to solve by choosing reasonable
boundary condition that $\varphi \to 0$ as $r_0 \to \infty $(Of
course, in de Sitter space there is a horizon such that $r_0$
cannot really go to $\infty$. However, since the horizon radius
$R$ is very large, one can take the horizon as $\infty$.). We
obtain
\be
\varphi=\frac{q}{4\pi r_{0}}e^{-\sqrt{3K}\,\displaystyle{r}_{0}}
\ee
This electric potential damps a little faster than in Lorentz
invariant electrodynamics. The electric field strength $\vec{E}$
at the observed point is
\be
\vec{E}=-q\frac{\vec{r}_{0}}{4\pi
r_{0}^{3}}e^{-\sqrt{3K}\,\displaystyle{r}_{0}}+q\frac{\sqrt{3K}\vec{r}_{0}}{4\pi
r_{0}^{2}}e^{-\sqrt{3K}\,\displaystyle{r}_{0}}.
\ee
This formalism is obviously different from the Coulomb Theorem.
Though the modification is very small, the electrostatic field
strength of a point charge in our LV electrodynamics model does
not exactly decay as $r^{-2}$. There is an another exponential
damping factor in the potential, which makes the potential looks
like a Yukawa one. This effect becomes important in the region of
far field and it may affect the large-scale universal observation.
However, since $K=\frac{1}{R^{2}}=\frac{\Lambda}{3}$ and $R$ could
be a very large distance parameter, say the 'radius of universe
horizon', the effect the exponential damping factor can be
negligible in the existing experiments.

Next we turn to focus our attention on magnetostatic field in
static de Sitter space-time. The simplest and also the most
fundamental case is the magnetic field of a small circle electric
current. We set the center of the small circle current (with the
electric current strength $I$ and the radius $a$) at the point
$\vec r_{0}$, and the observer is at the origin as the case of
electrostatic field mentioned above. This case, however, is a
little different from the electrostatic field, because the gauge
condition we are apt to select is the de Sitter gauge condition
Eq. (26). Using this gauge condition, one can arrive at
\be
\stackrel{\sim}\nabla\cdot\vec A=\frac{K}{\sqrt
\sigma}\vec{r}\cdot\vec{A}.
\ee
According to the electromagnetic field equations in static de
Sitter space-time, the differential equation of $\vec{A}$ can be
written as
\be
\stackrel{\sim}\nabla \times (\stackrel{\sim}\nabla \times
\vec{A})=\vec{\stackrel{\sim}j}(\vec{r})
\ee
here from Eq. (34)
\be
\vec{\stackrel{\sim}j}(\vec{r})=\sqrt{\sigma}\vec{j}(\vec{r}),\,\,
\vec{j}(\vec{r})=j_{r}\vartheta^{1}+j_{\theta}\vartheta^{2}+j_{\phi}\vartheta^{3}
\ee
is the conserved electric current in the local frame. However,
under this formalism, we don't know how to solve it. So we should
rewrite this equation in spherical coordinate as we did in the
case of electrostatic field of a point charge. To do this, one
should look back the essential meaning of an 'ordinary' 3-d vector
in a local frame, actually, a vector in $S^{3}$ space
$\vec{f}=f_{r}\vartheta^{1}+f_{\theta}\vartheta^{2}+f_{\phi}\vartheta^{3}$
can be corresponded to a vector in 3-d Euclidean space
$\vec{f}=\frac{1}{\sqrt{\sigma}}f_{r}\vec{e}_{r}+f_{\theta}\vec{e}_{\theta}+f_{\phi}\vec{e}_{\phi}$,
with this corresponding and define $\vec{A'}(\vec
r)=\frac{1}{\sqrt{\sigma}}A_{r}\vec{e}_{r}+A_{\theta}\vec{e}_{\theta}+A_{\phi}\vec{e}_{\phi}$,
then one can obtain the spherical coordinate formalism of the
gauge condition Eq. (40),
\be
\nabla\cdot \vec{A}'=4KrA'_{r}+Kr^{2}A'_{r,\,\,r}
\ee
Then the equations of the components of $\vec A'$ can be derived
directly from the Eq. (41)
\bea
-(\nabla^{2}\vec{A}')_{r}+ 4K A'_{r}+ 6K r
A'_{r},\,r+Kr^{2}A'_{r,\, r\,
,r}=\frac{1}{\sqrt{\sigma}}j_{r}\\
-(\nabla^{2}\vec{A}')_{\theta}+[K(r(rA'_{\theta})_{,\,r})_{,\,
r}+3KA'_{r,\,\theta}]=j_{\theta}\\
-(\nabla^{2}\vec{A}')_{\phi} +[ K(r(rA'_{\phi})_{,\,r})_{,\,
r}+\frac{3}{\sin{\theta}}KA'_{r,\,\phi}]=j_{\phi}.
\eea
Here, however, we have no reason to say that the vector
$(\frac{1}{\sqrt\sigma}j_{r}, j_{\theta},j_{\phi})$ is just a
conservation current density in spherical coordinates. Actually,
in static de Sitter space-time, the conservation current density
vector must be defined from the Eq. (34). Therefore we can obtain
the current strength $I$ through a certain cross section $S'$ in
spherical coordinates
\bea
I=\int {(\sqrt \sigma  \vec j)}  \cdot d\vec S' = \int {\sqrt
\sigma  } j_r rd\theta ' \wedge r'\sin \theta 'd\phi ' +  \\
\nonumber \int {j_\theta  } dr' \wedge r'\sin \theta 'd\phi ' +
\int {j_\varphi  } dr' \wedge r'd\theta '.
\eea
So the conservation current density vector in spherical
coordinates is $\vec {j}''(r)=\sqrt\sigma
j_{r}\vec{e}_{r}+j_{\theta}\vec{e}_{\theta}+j_{\phi}\vec{e}_{\phi}$.
Under this definition, Eqs. (44) should be multiplied a factor
$\sigma$. In the limit of $r \to 0$, one can prove that a
symmetric solution of Eqs.(44)-(46) is a solution of a vector
equation as follows. It is easy to show that the solution of
Eqs.(44)-(46) can be obtained by Bio-Savart Theorem for a small
circle current placed at the origin in usual Lorentz invariant
electrodynamics by setting $K$ to 0. Then pulling the source back
to $\vec r_{0}$, one can obtain $A'_{r},A'_{\theta},A'_{\phi}$
with this solution. Suppose that the solution in the spherical
frame with origin at the center of the circle is
$A'_{r'},A'_{\theta'},A'_{\phi'}$, it is easy to show that the
only non-vanishing component is $A'_{\phi'}$ and
$\vec{A'}(0)=A'_{\phi'}\vec{e}_{\phi'} =-A'_{\phi
'}\sin{\theta}\sin{(\phi - \phi_{0})}\vec{e}_{r} -A'_{\phi
'}\cos{\theta}\sin{(\phi - \phi_{0})}\vec{e}_{\theta}-A'_{\phi
'}\cos{(\phi - \phi_{0})}\vec{e}_{\phi}$, where $(r_0, \theta_0,
\phi_0)$ is the spherical coordinates of $\vec{r_0}$. Substituting
this solution to $K$ terms in Eqs.(44)-(46) and noting that $
\sigma \to 1$ as $r \to 0$, one can obtain the vector equation as
follows
\be
-\nabla^{2}\vec{A'}(\vec{r})+4K\vec{A'}(\vec{r})=\vec{j'}(\vec{r})
\ee
This equation is invariant under a translation in parameter
$\textbf{x}$ space, so we can build a spherical coordinate frame
$(r',\theta ',\phi ')$ with origin at the point $r_{0}$.  It is
easy to show that the only non-vanishing component of the solution
is still $A'_{\phi '}$ and $\vec{A'}(0)=A'_{\phi'}\vec{e}_{\phi'}
=-A'_{\phi '}\sin{\theta}\sin{(\phi - \phi_{0})}\vec{e}_{r}
-A'_{\phi '}\cos{\theta}\sin{(\phi -
\phi_{0})}\vec{e}_{\theta}-A'_{\phi '}\cos{(\phi -
\phi_{0})}\vec{e}_{\phi}$ also holds, where
\be
\vec{A'}=A'_{\phi '} \vec {e}_{\phi
'}=\frac{Ia}{4\pi}\oint^{2\pi}_{0}\frac{\cos{\varphi}
d\varphi}{\sqrt{r_{0}^{2}+a^{2}-2r_{0}a\sin{\theta
'}\cos{\varphi}}}
e^{-\sqrt{4K}\sqrt{r_{0}^{2}+a^{2}-2r_{0}a\sin{\theta
'}\cos{\varphi}}}\vec {e}_{\phi '}
\ee
In the cases $2r_{0}a\sin{\theta '}\ll r_{0}^{2}+a^{2}$, namely in
the region of far field ($r_{0}\gg a$), and $r_{0}\sin{\theta
'}\ll a$, the so called region of adaxial field, the above
integral can be approximatively calculated to 3-order
\bea
A'_{\phi '}=\frac{Ia}{4\pi}\int d\varphi
\cos{\varphi}[\mathcal{P}\frac{r_{0}a\sin{\theta
'}\cos{\varphi}}{(r_{0}^{2}+a^{2})^{3/2}}+\mathcal{N}\frac{r_{0}^{3}a^{3}\sin^{3}{\theta
'}\cos^{3}{\varphi}}{(r_{0}^{2}+a^{2})^{7/2}}]\\ \nonumber
=\frac{Ia}{4\pi}[\mathcal{P}\frac{r_{0}a\sin{\theta
'}}{(r_{0}^{2}+a^{2})^{3/2}}+\frac{3}{4}\mathcal{N}\frac{r_{0}^{3}a^{3}\sin^{3}{\theta
'}}{(r_{0}^{2}+a^{2})^{7/2}}],
\eea
where
\bea
\mathcal{P}=\frac{e^{-\sqrt{4K(r_{0}^{2}+a^{2})}}}{2}(1+\sqrt{4K(r_{0}^{2}+a^{2})})\nonumber
\eea
and
\be
\mathcal{N}=\frac{e^{-\sqrt{4K(r_{0}^{2}+a^{2})}}}{48}\{15+15\sqrt{4K(r_{0}^{2}+a^{2})}
\\ \nonumber
+24K(r_{0}^{2}+a^{2})+[\sqrt{4K(r_{0}^{2}+a^{2})}\,\,]^{3}\}.
\nonumber
\ee
Pulling the origin back to the field point by setting $\theta
'=\pi -\theta_{0},\phi '=\phi _{0}-\pi$ then the magnetic
potential $\vec A$ (at the origin) of the circle electric current
can be written as
\bea
\vec
A(0)=\vec{A'}(0)=A'_{r}\vec{e}_{r}+A'_{\theta}\vec{e}_{\theta}+A'_{\phi}\vec{e}_{\phi}
=-A'_{\phi '}\sin{\theta}\sin{(\phi -
\phi_{0})}\vec{e}_{r}\\
\nonumber -A'_{\phi '}\cos{\theta}\sin{(\phi -
\phi_{0})}\vec{e}_{\theta}-A'_{\phi '}\cos{(\phi -
\phi_{0})}\vec{e}_{\phi}.
\eea
This solution shows that the vector potential of the circle
current is also a damping potential, it decays a little faster
than that in Lorentz invariant electrodynamics as the case of
electrostatic field of a point charge mentioned above. So it is
reasonable to say that the magnetic field strength also have a
damping factor. In the region of the far field, this damping can
not be ignored. In this LV electrodynamics approach, at least on
very large scale observation, the LV effect become important and
the observation data should be reconsidered because of the damping
factor.

In conclusion, we set up an effective low energy LV classical
electrodynamic model in Minkowski space-time by the covariant
fomulation of electrodynamic in static de Sitter space-time. We
define the observable in the model as the vierbein decomposition
components of physical tensors. The electromagnetic field
equations are obtained in this formalism and its deviation from
ones in Lorentz invariant theory is showed. Furthermore, we
investigate the energy-momentum conservation law in this LV model.
As an application of the LV electromagnetic equations, we studied
two basic and simple cases which might be responsible for possible
observation confirmation: the electrostatic field of a point
charge and the magnetostatic field of a circle electric current.
We find that in both case there is an analogous damping factor in
the potential function, this can be regarded as a side face of LV
effect and may be important on large scale observations.

%%%%%%%%%%%%%%%%%%%%%%%%%%%%%%%%%%%%%%%%%%%%%%%%%%%%%%%%%%%%%%%%%%

\end{document}